# Temperature driven spin-zero effect in TaAs$_2$

Md Shahin Alam [1], P.K. Tanwar [1], Krzysztof Dybko [1,2], Ashutosh S. Wadge [1], Przemysław Iwanowski [1,2], Andrzej Wiśniewski [1,2] and Marcin Matusiak [1,3,*]

1. *International Research Centre MagTop, Institute of Physics, Polish Academy of Sciences, Aleja Lotnikow 32/46, PL-02668 Warsaw, Poland*
2. *Institute of Physics, Polish Academy of Sciences, Aleja Lotników 32/46, PL-02668 Warsaw, Poland*
3. *Institute of Low Temperature and Structure Research, Polish Academy of Sciences, ul. Okólna 2, 50-422 Wrocław, Poland*

Electrical and thermo-electrical transport properties of the semimetal TaAs$_2$ were measured in a magnetic field applied along [-2 0 1] direction. The resulting field dependences of the resistivity as well as the Hall, Seebeck and Nernst coefficient below $T \approx 100$ K can be satisfactorily described within the two-band model consisting of electron and hole pockets. At low temperature all the measured coefficients exhibit significant contribution from quantum oscillations. The fast Fourier transform (FFT) of the oscillatory Nernst signal shows two fundamental frequencies, $F_\alpha = 105$ T and $F_\beta = 221$ T, and the second harmonic of the latter ($F_{2\beta} = 442$ T). The ratio between FFT amplitudes of $F_\beta$ and $F_{2\beta}$ changes with temperature in an unusual way, indicating that we observe a spin-zero effect caused by temperature change. This is likely related to the substantial temperature dependence of the Landé g-factor, which in turn can result from the non-parabolic energy dispersion or temperature evolution of the spin-orbit coupling.

1. Introduction

The realisation of the role, which the topological structure plays in determining macroscopic properties of a system, led to one of the biggest breakthroughs in the modern solid state physics [1-3]. Subsequent research activity resulted in the discovery of a large variety of different types of topologically non-trivial phases [4-16]. $TaAs_2$ is an interesting topological semimetal [17,18] that was suggested to have trivial strong topological index along with three non-trivial weak indices [19]. Strong and weak surface states are, for example, immune or sensitive to disorder, respectively [20]. It is also expected that the Zeeman effect in $TaAs_2$ induces additional Berry curvature, which makes it a type-II Weyl topological semimetal in the magnetic field [21] or change the electronic structure from trivial to non-trivial (and vice versa) depending on the specific directions of the magnetic field [22].

In this work, we investigate magneto-thermo-electrical properties of $TaAs_2$ below $T \approx 100$ K. The field dependences of the transport coefficients in this range can be satisfyingly modelled with a semi-classical multiband approach, but in the Nernst effect we observe an unusual temperature evolution of quantum oscillations. Namely, the relation between amplitudes of the 221 T fundamental frequency and its second harmonic indicates occurrence of the spin-zero effect in $TaAs_2$. Remarkably, the phenomenon arises as a result of changes in temperature, and not the angle at which the magnetic field is applied. This opens the possibility that topological attributes of $TaAs_2$ are temperature dependent.

2. Material and methods

Single crystals of $TaAs_2$ were grown by the two-stage chemical vapour transport. Polycrystalline $TaAs_2$ was synthesized by a direct reaction of Ta foil (ZR industrial Ltd, 99.99%) and As (PPM Pure Metals, 99.999995%). Reagents were placed inside the evacuated quartz tube at 990 °C for 19 days. After synthesis, polycrystalline $TaAs_2$ was pressed into pellets, which were loaded into quartz tube with Iodine (POCH, 99.8%), sealed under the vacuum. The tube was then placed into a gradient zone of temperature 1025 °C (crystallization zone) and 956 °C (source zone) for 23 days, after which the furnace was cooled down to room temperature at 100 °C/h. X-ray powder diffraction confirmed the monoclinic unit cell and C2/m (No. 12) space group of the studied crystals. Surface quality and quantitative chemical composition was verified using energy dispersive X-ray spectroscopy (EDX) system QUANTAX 400 Bruker coupled with the Zeiss Auriga field emission (Schottky type) scanning electron microscope (FESEM) [23].

For the transport measurements a single crystal was cut to a plate with dimensions 1.7 x 1.3 x 0.3 $mm^3$. The crystallographic b-axis was oriented along the longest side of the sample and electrical and thermal current was applied along this [0 1 0] direction. The magnetic field was applied perpendicularly to the naturally occurring (-2 0 1) plane.

For the resistivity measurements, the electrical contacts were arranged in the Hall bar geometry. During the Seebeck ($S_{xx}$) and Nernst ($S_{xy}$) signal measurements, the sample was clamped between two phosphor bronze blocks, which had two Cernox thermometers and resistive heaters attached to them. The magnetic field sweeps were performed in magnetic fields from -14.5 to +14.5 T in order to extract the voltage components that were odd and even in $B$. For the temperature ramps in $B = 0$ T, the "heater on-off" method was used, while for the field sweeps, the heater was permanently turned on. The temperatures were measured with Cernoxes thermometers that were supposed to exhibit minimal magnetoresistance. In the range they were used (i.e. above $T = 5$ K and up to $B = 14.5$ T), the relative error related to the magnetic field influence was expected to be smaller than ~2% [24,25].

## 3. Results

Exemplary field dependences of the electrical resistivity ($\rho_{xx}$) and Hall resistivity ($\rho_{xy}$) for TaAs$_2$ are presented in Fig. 1. These are the results of symmetrisation and anti-symmetrisation of the data collected for the positive and negative magnetic field. The temperature dependence of $\rho_{xx}$ is shown in inset of Fig. 1a. The value of the residual resistivity ratio (RRR = $\rho_{xx}$(300K)/$\rho_{xx}$(3K)) is about 95, which indicates good quality of the sample, and it is comparable with previously reported values (RRR ≈ 85 [26], 100 [18], 120 [27]). One of the hallmarks of TaAs$_2$ is its extremely large and non-saturating magnetoresistance [18,19,26-29]. In our case $\rho_{xx}$ in the low temperature limit increases almost 35 000 – fold between $B = 0$ and 14.5 T. Such a large magnetoresistance could be a consequence of the fact that TaAs$_2$ appears to be a nearly compensated semimetal in agreement with the density-functional theory calculations [17], alternatively, this might also stem from its non-trivial topology [19,22].

The slope of the Hall resistivity dependences, presented in Fig. 1b, exhibits a similar, strong dependence on the magnetic field, i.e. $\rho_{xy}/B$ is very small in low field, but becomes large and negative at high field, which suggests dominating role of electrons at high field. For example, at $T = 1.7$ K calculated from the linear fit in the range 0.035 T $< B <$ 0.5 T $\rho_{xy}/B \approx$ -6 ± 2 x 10$^{-9}$ Ω m / T (= m$^3$ / C) and $\rho_{xy}/B \approx$ - 4.5 x 10$^{-7}$ m$^3$ / C for $B = 14.5$ T.

The $\rho_{xx}(B)$ and $\rho_{xy}(B)$ dependences can be reasonably well modelled using semi-classical two band approximation [30]:

$$\rho_{xx}(B) = \frac{1}{e} \frac{(n_e\mu_e+n_h\mu_h)+(n_h\mu_e+n_e\mu_h)\mu_e\mu_h B^2}{(n_e\mu_e+n_h\mu_h)^2+[(n_h-n_e)\mu_e\mu_h]^2 B^2}, \quad (1)$$

$$\rho_{xy}(B) = \frac{B}{e} \frac{(n_h\mu_h^2-n_e\mu_e^2)+(n_h-n_e)\mu_e^2\mu_h^2 B^2}{(n_e\mu_e+n_h\mu_h)^2+[(n_h-n_e)\mu_e\mu_h]^2 B^2}, \quad (2)$$

where $n$ is concentration, $\mu$ mobility, $e$ elementary charge, whereas $e$ and $h$ indexes denote the electron and hole contributions, respectively. Fits of both $\rho_{xx}(B)$ and $\rho_{xy}(B)$ for a given temperature were done simultaneously and are shown in Fig. 1. In the low temperature limit they give values of concentrations of both electrons and holes around $10^{25}$ m$^{-3}$ ($n_e$ = 1.1 x $10^{25}$ m$^{-3}$ and $n_h$ = 1.0 x $10^{25}$ m$^{-3}$ at $T$ = 1.7 K, respectively) and mobilities around 0.5 m$^2$ / V s ($\mu_e$ = 0.52 m$^2$ / V s and $\mu_h$ = 0.45 m$^2$ / V s at $T$ = 1.7 K). These are in the same order of magnitude as previously reported numbers [23].

At low temperature and high field both $\rho_{xx}$ and $\rho_{xy}$ exhibit strong oscillatory component as a result of the Landau level quantization. Inset in Fig. 1b presents exemplary ($T$ = 5.4 and 19.8 K) fast Fourier transforms (FFT) of the resistivity calculated for the field range 8 – 14.4 T with a subtracted background in a form of the 3$^{rd}$ order polynomial. For the unevenly sampled signal we used the algorithm based on the Lomb normalized periodogram. We are able to confirm presence of two fundamental oscillations: $F_{\rho\alpha}$ = 122 T, $F_{\rho\beta}$ = 210 T, and the second harmonic of the latter: $F_{2\rho\beta}$ = 420 T. At $T$ = 5.4 K we observe double-peaks structures that might be due to additional orbits present in the irregular shaped electron/hole pockets [17]. Similar behaviour was, for instance, reported for NbP [31] or, due to a small misalignment of the magnetic field, in Sb [32,33].

Analogously to $\rho_{xx}(B)$ and $\rho_{xy}(B)$, the Seebeck (Fig. 2a) and Nernst (Fig. 2b) signals also show strong dependence on the magnetic field. The absolute value of $S_{xy}$ at $B$ = 14.5 T shows some non-monotonic temperature dependence, which would be removed if it was divided by temperature to account for entropy changes. The data can be fitted with the conventional semi-classical equations [34,35]:

$$S_{xx}(B) = S_{xx}^0 \frac{1}{1+(\mu B)^2} + S_{xx}^\infty \frac{(\mu B)^2}{1+(\mu B)^2}, \qquad (3)$$

$$S_{xy}(B) = S_{xy}^0 \frac{\mu B}{1+(\mu B)^2}, \qquad (4)$$

where $S_{xx}^0$ and $S_{xx}^\infty$ are values of the thermoelectric power in the limit of zero and infinite field, respectively, $S_{xy}^0$ is the amplitude of the Nernst signal in the zero field limit, $\mu = \sqrt{\mu_e \mu_h}$ is the mean mobility [30]. There are some deviations of the fitting lines from the actual data for the Nernst signal above $T \approx 50$ K, but overall, $S_{xx}(B)$ and $S_{xy}(B)$ can be well approximated using Eqns. 3 and 4. However, the resulting $\mu$ values are about one orders of magnitude smaller than ones deducted from electrical transport measurements (despite they are consistent between $S_{xx}(B)$ and $S_{xy}(B)$). We conclude that despite Eqns. 3 and 4, which were formulated for a single band conductor [34], can satisfactorily reproduce the field dependence of the thermoelectrical coefficients in a multiband semimetal, the resulting mobilities are not best approximation of the actual values. Probably, it is due to not complete compensation of electron and hole carrier densities in TaAs$_2$ (at $T = 1.7$ K the electrical conductivity of electrons in TaAs$_2$ is about 27% higher than the electrical conductivity of holes).

A multiband character of TaAs$_2$ can also explain the field dependence of the thermoelectric power, which similarly to $\rho_{xy}/B$ is very small when $B \rightarrow 0$ T at $T = 5.4$ K, and becomes substantially negative in field, reaching $S_{xx}(14.5\,\text{T}) \approx -12$ μV / K. Inset in Fig. 2a, showing the temperature dependence of $S_{xx}$ in the zero magnetic field, indicates that almost perfect compensation of $S_{xx}$ in zero field occurs below $T \approx 80$ K. At low temperature and high magnetic field, a large oscillatory component appears in both $S_{xx}(B)$ and $S_{xy}(B)$. Inset in Fig. 2b presents FFT of the latter at two temperatures: $T = 5.4$ and 35 K (which were calculated for the field range 8 – 14.5 T with a subtracted 3$^{rd}$ order polynomial as a background), where, again, two fundamental frequencies are present: $F_{\nu\alpha} = 105$ T, $F_{\nu\beta} = 221$ T. They are slightly different than ones estimated from the Shubnikov - de Haas effect, most likely due to small misalignment of the crystal against the magnetic field in two separate experiments. Here, we observe the second harmonic of the β oscillations: $F_{2\nu\beta} = 442$ T, which in FFT is accompanied by the somewhat smaller peak at $F_{2\nu\beta}' = 413$ T. We believe it is the second harmonic of the twin β peak at $F_{\nu\beta}' = 206.5$ T. With our resolution we are not able to distinguish it from larger $F_{\nu\beta} = 221$ T, but we notice a beating pattern in the oscillatory signal of β band filtered out of other frequencies (see Fig. S4 in the Supplementary Material (SM)). This indicates the presence of two close frequencies. We expect that this is a consequence of the aforementioned shift in angle, which for an irregular electron/hole pocket (like ones present in the calculated electronic structure of TaAs$_2$ [17]) can create another extremal Fermi surface area. As a general note, we would like to note that the signal to noise ratio from the Nernst data was the best among measured quantities, thus we focus mostly on analysing these results.

## 4. Discussion

The most striking result of our work is presented in Fig. 3, where the oscillatory components of the Nernst signal at two different temperatures are compared. Namely, it turns out that $F_{\nu\beta}$ dominating at $T = 11.2$ K is almost entirely replaced by its second harmonic, $F_{2\nu\beta}$, at $T = 24.6$ K. In regular metals the higher harmonics of the quantum oscillations disappear faster with the temperature than the fundamental frequency. The amplitude of the $p$-th harmonic of the oscillations is reduced by the factor $R_T = \frac{\lambda p T}{\sinh(\lambda p T)}$ [36], where $\lambda = \frac{2\pi^2 k_B m^*}{e \hbar B}$ ($k_B$, $m^*$ and $\hbar$ is the Boltzmann constant, effective mass and reduced Planck constant, respectively). The upper inset in Fig. 4 presents changes of the relative amplitude of the first and second harmonic of the β oscillations for three representative temperatures. At high temperature the amplitude of $F_{\nu\beta}$ is largely suppressed, while $F_{2\nu\beta}$ is clearly visible and the reduction factor in this region is anomalously small (i.e. smaller than 1). The ratio of the 1$^{st}$ to 2$^{nd}$ harmonics amplitude $A_1^\nu / A_2^\nu = 0.13$ at $T = 24.5$ K, whereas at $T = 34.7$ K it increases slightly ($A_1^\nu / A_2^\nu = 0.28$), perhaps departing from zero at the temperature between 24.5 and 34.7 K. In Fig. 4 we present the entire temperature dependence of $A_1^\nu / A_2^\nu$ calculated from the Nernst signal, which seems to be confirmed by $A_1^\rho / A_2^\rho$ ($T$) deducted from the resistivity shown in the lower inset of the same figure (separated temperature dependences of amplitudes are shown in Figs. S2 and S3 in SM). In the latter case, the signal to noise ratio is somewhat lower than one from the Nernst effect, but overall appearance of $A_1 / A_2$ remains similar to one from the Fig. 4 main panel.

The effect of suppressing the fundamental amplitude with enhancement of the second harmonic is a consequence of Zeeman splitting and it is called the spin-zero effect [36]. This happens when:

$$\cos\left(\frac{\pi}{2} g \frac{m^*}{m_e}\right) = 0, \qquad (5)$$

and hence that $g = (2r + 1)/\frac{m^*}{m_e}$, where $r$ is any integer, and $g$ is the Landé $g$-factor.

The spin-zero effect is normally observed when the condition described by the Eqn. 5 is met for a particular orientation of applied magnetic field. Its occurrence was reported in regular metals like copper [37,38], or gold and silver [39]. Recently, such a phenomenon was also observed in the Weyl semimetal $WTe_2$ [40] and the Dirac semimetal $ZrTe_5$ [41]. Despite the relativistic energy dispersion in these topological materials, the same Eqn. 5 must be met for the spin-zero effect to occur [41]. However, it was shown that the additional Zeeman-induced Berry curvature shifts the spin-zero angle in $ZrTe_5$ [22]. Remarkably, S. Sun et al. [22] predicted that the spin-zero effect is also expected to occur in $TaAs_2$ and that it should be affected by the Berry curvature similarly to $ZrTe_5$. In fact, the angular dependence of the Shubnikov – de Haas effect at $T = 1.6$ K shows that the fundamental frequency amplitude of β oscillations nearly vanishes for the field applied around 30 degree from [-2 0 1] [23].

Since the spin-zero effect emerges due to the phase shift between the oscillations coming from the spin-up and spin-down electrons, it can be expected that the phase of the fundamental frequency will be inverted at the spin-zero condition. In fact, such a phenomenon was reported at the spin-zero angle in $ZrTe_5$ [22]. Notably, we also observe the phase inversion of the β oscillations around at $T = 24.5$ K. Figure 5 shows a comparison of the Nernst oscillations at the temperatures below ($T = 11.2$ K) and above ($T = 35$ K) the spin-zero effect, where the β frequency, filtered out of the latter, appears to be in antiphase with the β oscillation at low temperature.

We would like to stress that our discovery of the spin-zero effect in TaAs$_2$ is unusual in terms of condition that triggers the phenomenon. Namely, it is not caused by changes of the angle, but by changes of the temperature. As indicated by Eqn. 5 this can only happen due to variation of the g-factor or effective mass. Since the frequencies of the oscillations (hence areas of the Fermi surface extremal cross sections) do not change much with temperature, we believe that $T$ - dependence of the g-factor may be the main cause of the effect in TaAs$_2$.

In order to estimate the *g*-factor at $T$ = 5.4 K, we used the oscillatory part of the Nernst signal presented in Fig. 6, where peaks for $B$ > 11 T begin to split due to the Zeeman effect. If the spin degeneracy is lifted, the Landau level index plot should be prepared separately for spin-up and spin-down Landau levels (see inset in Fig. 6): $n = \frac{F}{B} + \delta + \frac{1}{2}\varphi$, where *n* is the Landau level index, δ is the phase shift related to topology and dimensionality of a band, and $\varphi$ is the phase difference between up ($\varphi = \frac{gm^*}{2m_e}$) and down ($\varphi = -\frac{gm^*}{2m_e}$) spins [42,43]. This procedure gives the *g*-factor at $T$ = 5.4 K equal to $g_0$ = 5.8.

Using the effective mass of the β band $m^*$ = 0.35 $m_e$, estimated from the Shubnikov – de Haas effect (see Fig. S5 in SM), the spin-zero effect can be used to estimate the value of *g* at the temperature close to 25 K. The closest to $g_0$ values of the g-factor satisfying Eqn. 5 are $g$ = 2.9 for $r$ = 0 and $g$ = 8.6 for $r$ = 1. In either case, g at $T \approx$ 25 K is significantly different from $g_0$ = 5.8 at $T$ = 5.4 K.

Observations of a temperature dependent g-factor were already reported for several semiconductors, such as CdTe and GaAs [44-48]. In the latter, the temperature variation of *g* was ascribed to band's non-parabolicity and evolution of the energy gap [48]. While relative changes of the g-factor in semiconductors are rather modest, the significant variation of *g* with temperature was reported for the strongly correlated compound, SrIrO$_3$ [49]. In this case, temperature dependent spin-orbit coupling (SOC) was identified as an origin of the phenomenon. In this material, as in the case of the Dirac semimetal $Cd_3As_2$ [50], the temperature dependent SOC was suggested to be related to changes of the Rashba coefficient. However, based on the current studies we are unable to pinpoint what mechanism could drive changes in SOC in $TaAs_2$.

In general, SOC and Berry phase can be related [51], but the scenario where the g-factor changes due to SOC temperature variation is of particular interest in $TaAs_2$ as the topological attributes of this material are determined by the SOC strength [52]. Namely, according to calculations based on the density functional theory in the absence of SOC, $TaAs_2$ possess two types of bulk nodal lines. Turning on SOC opens a continuous bandgap in the energy spectrum and drives the system into a topological crystalline insulator. In other words, changes of SOC in $TaAs_2$ can affect topology of the electronic system, hence open a possibility of emergence of the temperature driven topological transition. Whether the evolution of the Seebeck and Nernst signal that we observe in $TaAs_2$ is somehow related to its topological properties, can be an exciting subject for further experimental and theoretical investigations.

## 5. Conclusions

The temperature dependences of the quantum oscillations in the Nernst effect and resistivity indicate that for the magnetic field parallel to [-2 0 1] and $T \approx 25$ K we observe the spin-zero effect in $TaAs_2$. The Landé g-factor estimated for this temperature is significantly different from that calculated for $T = 5.4$ K. Among possible origins of the temperature evolution of the g-factor are band's non-parabolicity, evolution of the energy gap or changes in the spin-orbit coupling. The latter is of particular interest because in $TaAs_2$ it can lead to modification of the electronic system's topology.

**Acknowledgments**

We would like to thank Carmine Autieri for helpful discussion.

This work was supported by the Foundation for Polish Science through the IRA Programme co-financed by EU within SG.

**Competing financial interests:**

The authors declare no competing financial interests.
**Data availability**

All of the relevant data that support the findings of this study are available from the corresponding author upon reasonable request.

**M. Matusiak ORCID iD**: 0000-0003-4480-9373
11

**Figures**

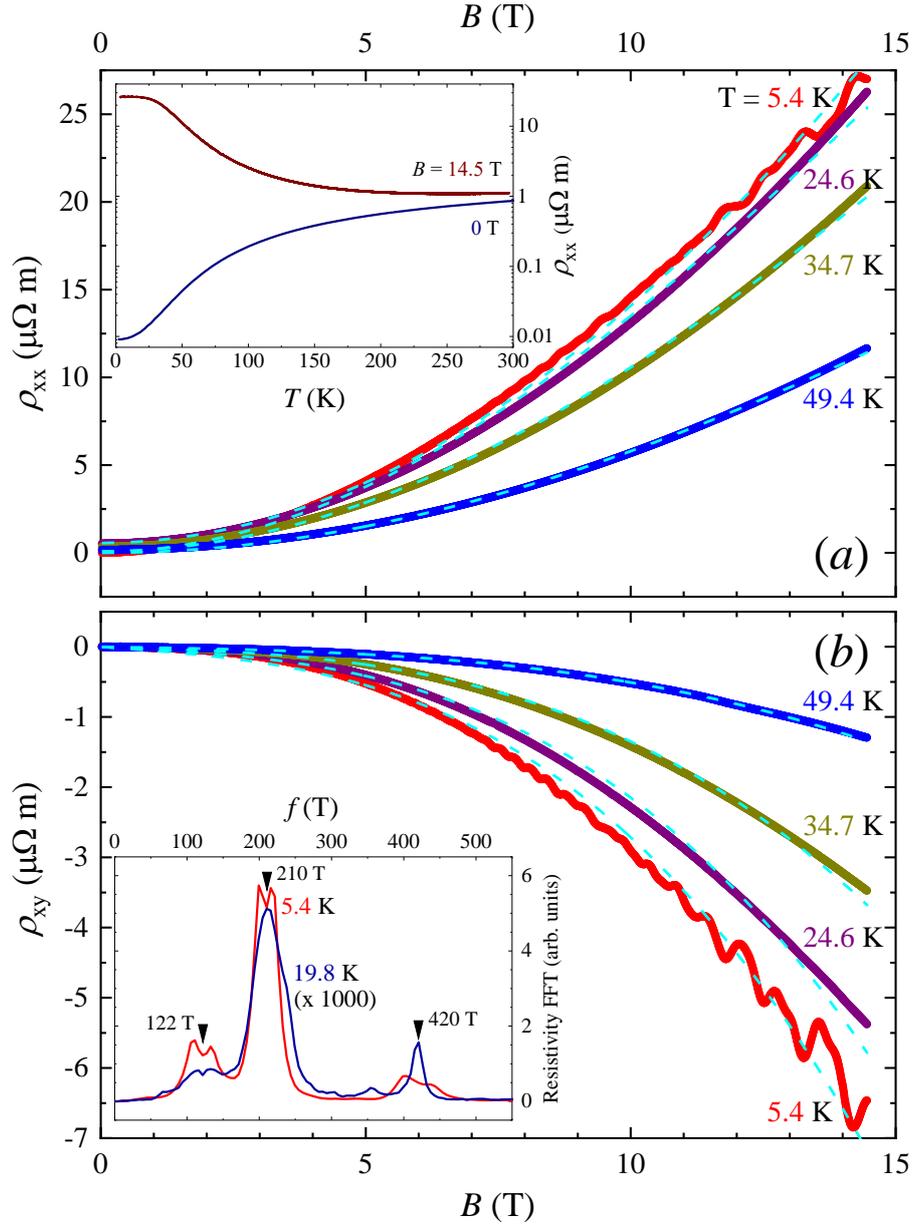

**Figure 1.** (Color online) Magnetic field dependence of the longitudinal resistivity (panel *a*) and Hall resistivity (panel *b*) for various temperatures with current direction parallel to [0 1 0] and ***B*** parallel to [-2 0 1]. The dashed lines in both panels show fits prepared within the semi-classical two band model (Eqns. 1 and 2).

Inset in panel *a* presents temperature dependence of $\rho_{xx}$ for 0 T and 14.5 T magnetic field, inset in panel *b* shows the fast Fourier transform spectrum of $\rho_{xx}$ at 5.4 K (red) and 19.8 K (blue).

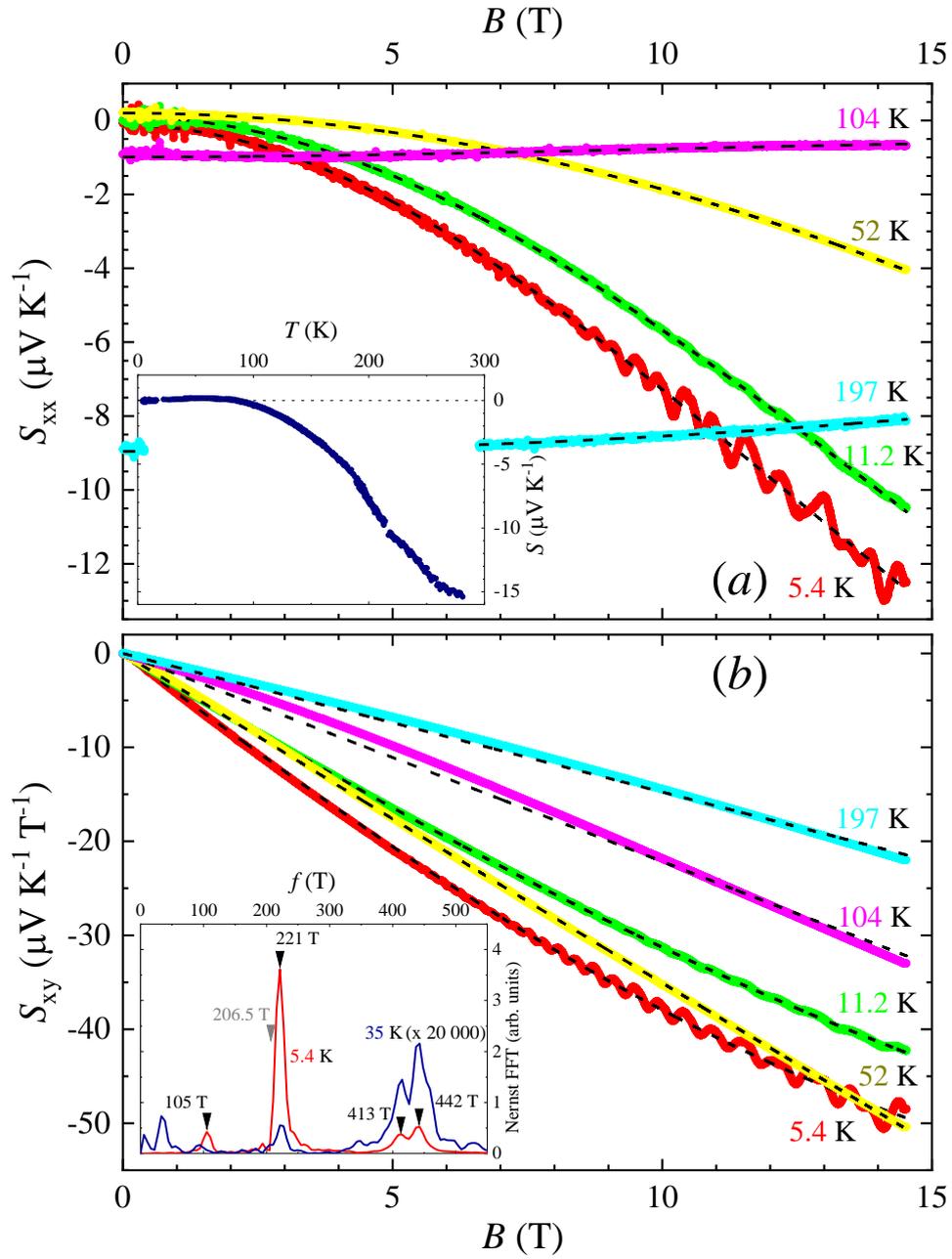

**Figure 2.** (Color online) Magnetic field dependence of the thermoelectric power (panel *a*) and Nernst coefficient (panel *b*) for various temperatures. The dashes lines in both panels show fits prepared using equations 3 and 4.

Inset in panel *a* presents temperature dependence of the Seebeck coefficient at zero magnetic field, inset in panel *b* shows FFT spectrum of the Nernst coefficient at 5.4 K (red) and 35 K (blue).

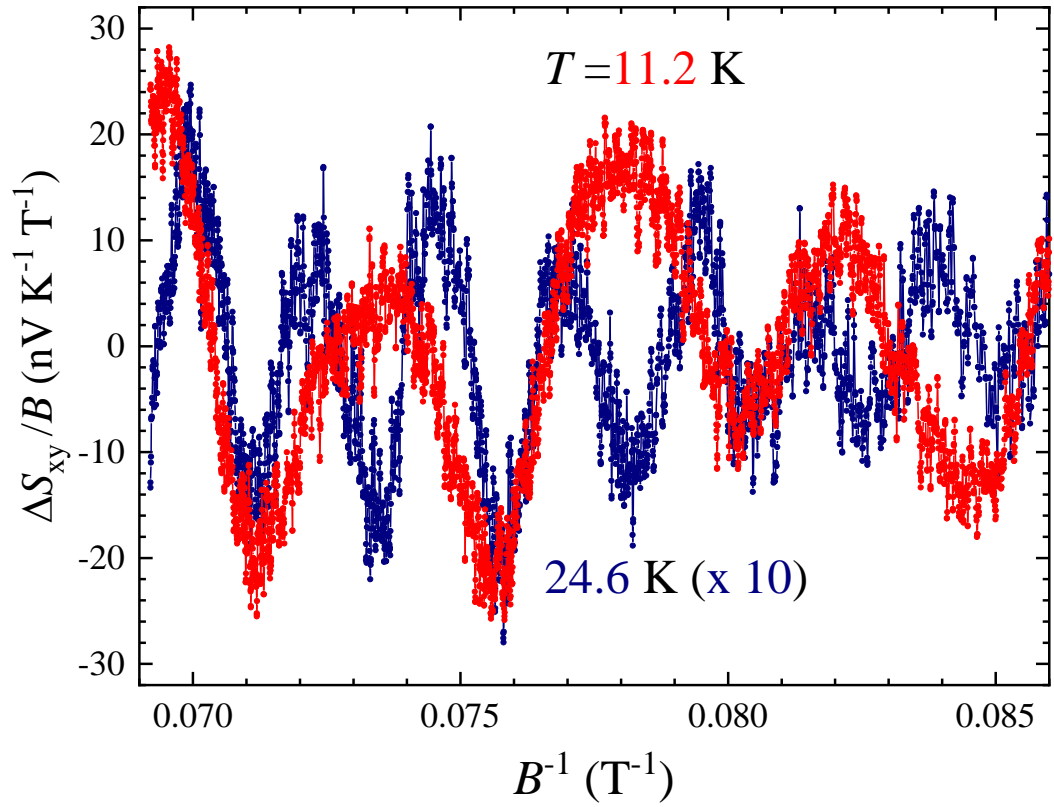

**Figure 3.** Comparison of the oscillatory component of the Nernst signal in TaAs$_2$ plotted versus inverse magnetic field for $T = 11.2$ K and 24.6 K. The results for the latter are multiplied by a factor of 10.

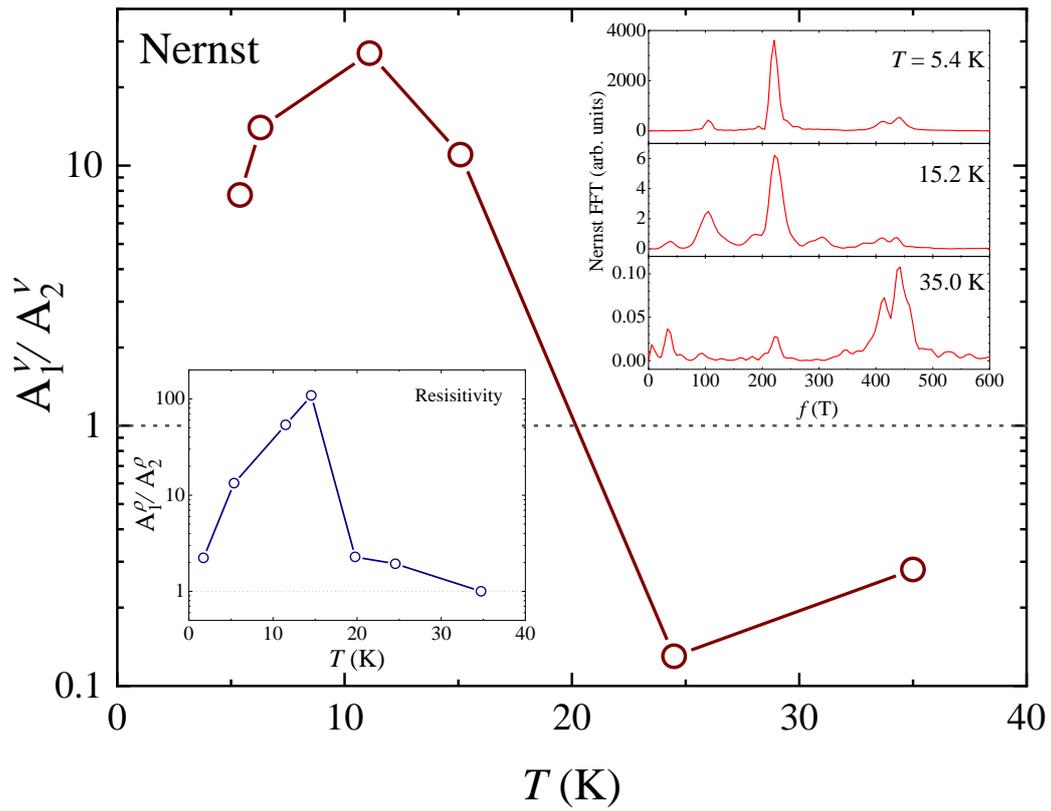

**Figure 4.** (Color online) Calculated from the Nernst signal temperature dependence of the first to second harmonic ratio for the β band. Upper inset shows evolution of Nernst FFT spectrum around β frequency and its second harmonic at $T = 5.4$ K, 15.2 K and 35 K. The lower inset presents the analogous amplitude ratio as the main panel, but estimated from the resistivity data.

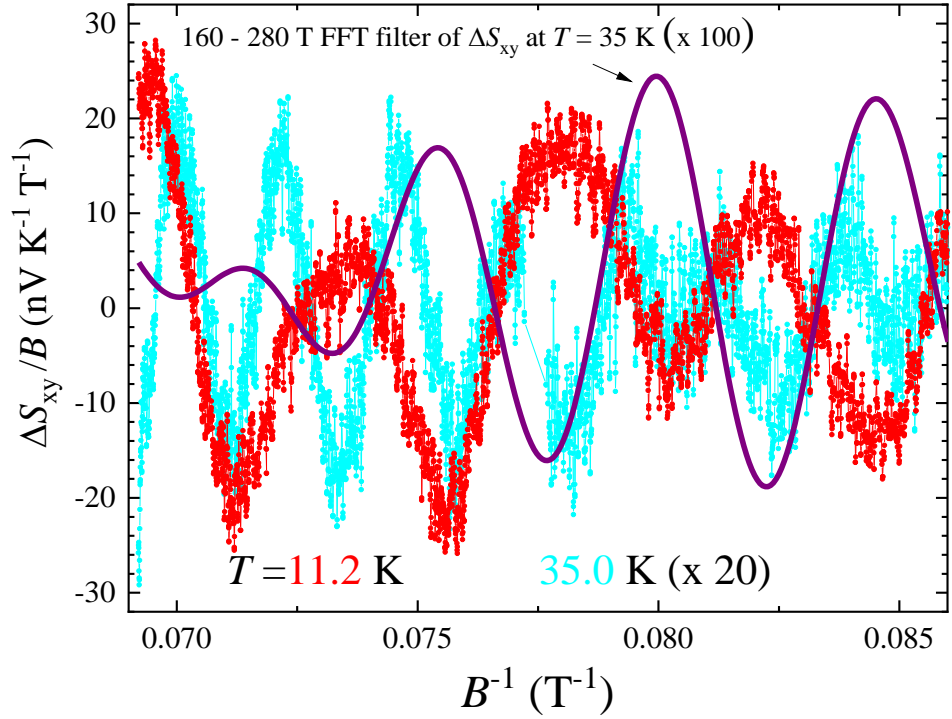

**Figure 5.** Comparison of the oscillatory component of the Nernst signal plotted versus inverse magnetic field for $T = 11.2$ K and 35 K. The latter is multiplied by a factor of 20. The solid line depicts the signal at $T = 35$ K filtered with the FFT band-pass filter for the 160 - 280 T frequency range, which turns out to be in antiphase with the fundamental β oscillation below $T \approx 25$ K.

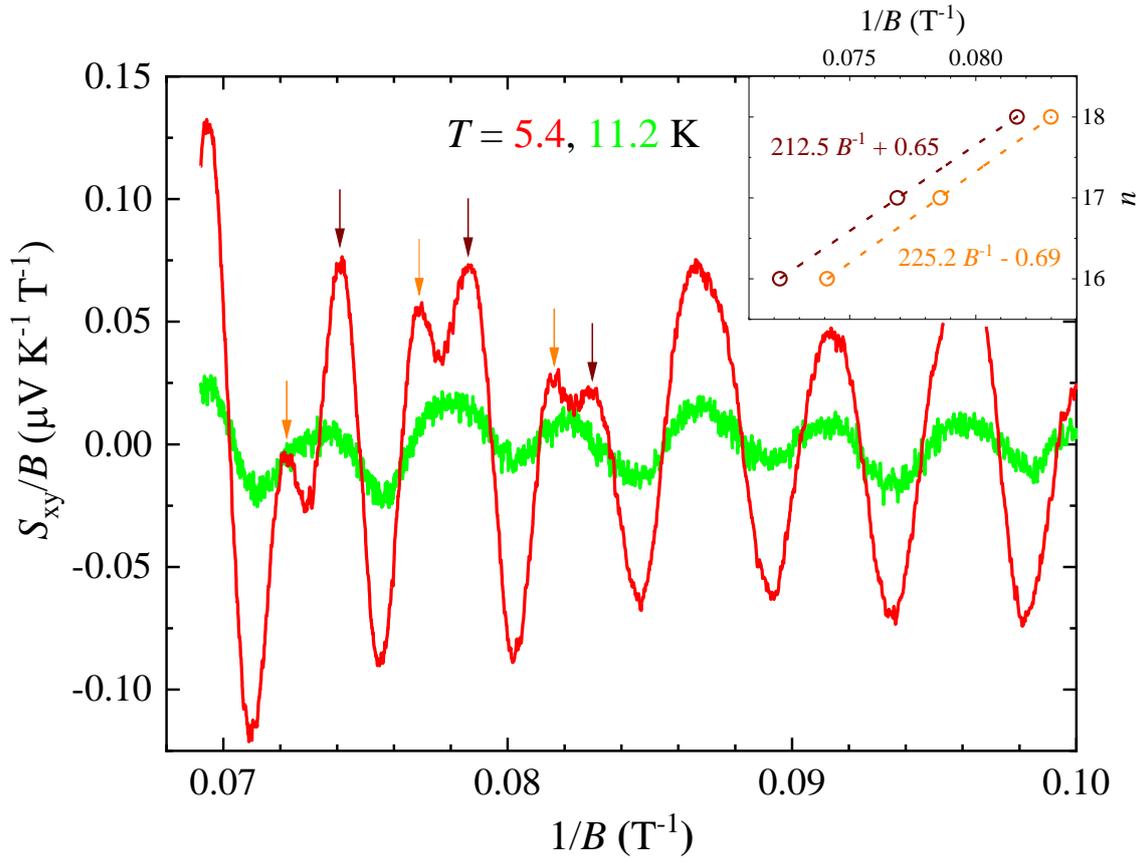

**Figure 6.** (Color online) Oscillatory part of the Nernst signal plotted versus $B^{-1}$ for $T = 5.4$ K (where peaks split at high magnetic field) and 11.2 K (where the split is absent). Inset presents Landau level fan diagrams for $T = 5.4$ K separated by the Zeeman splitting for the spin-up and spin-down plots.

# Supplementary material: Temperature driven spin-zero effect in TaAs$_2$

Md Shahin Alam [1], P.K. Tanwar [1], Krzysztof Dybko [1,2], Ashutosh S. Wadge [1], Przemysław Iwanowski [1,2], Andrzej Wiśniewski [1,2] and Marcin Matusiak [1,3,*]

1. International Research Centre MagTop, Institute of Physics, Polish Academy of Sciences, Aleja Lotnikow 32/46, PL-02668 Warsaw, Poland

2. Institute of Physics, Polish Academy of Sciences, Aleja Lotników 32/46, PL-02668 Warsaw, Poland

3. Institute of Low Temperature and Structure Research, Polish Academy of Sciences, ul. Okólna 2, 50-422 Wrocław, Poland

## A: Nernst signal oscillations in TaAs$_2$ and corresponding fast Fourier transform spectra

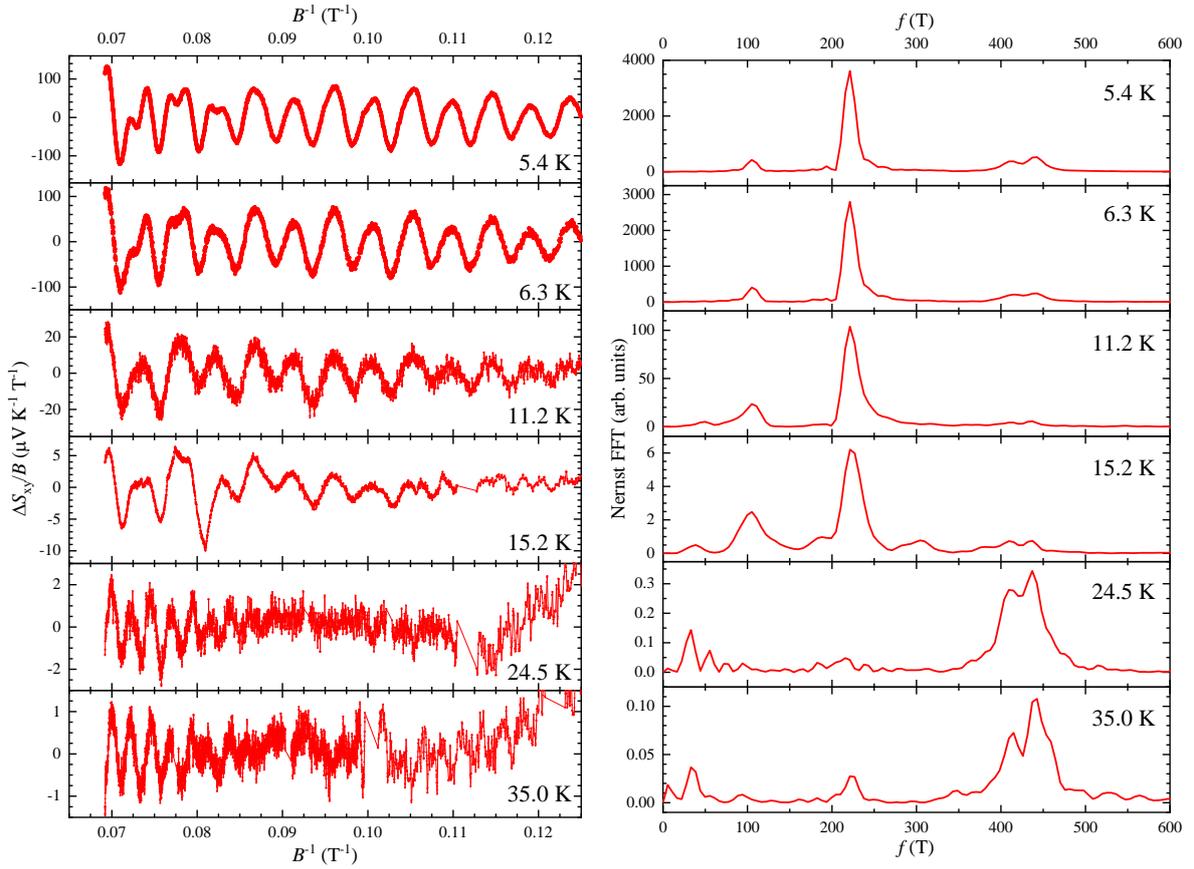

**Figure S1.** The oscillatory component of the Nernst coefficient in TaAs$_2$ plotted versus inverse magnetic field for various temperatures (left panel). The right panel presents respective FFT spectra.

**Table S1.** Parameters obtained from Lorentzian multiple peak fitting of the Nernst FFT spectra for β and 2β oscillations.

| $T$ (K) | 1st harmonic amplitude (arb. units) | 2nd harmonic amplitude (arb. units) | $A_1/A_2$ | 1st harmonic Frequency (T) | 2nd harmonic Frequency (T) |
|---|---|---|---|---|---|
| 5.4 | 3.81E-16 | 4.94E-17 | 7.7 | 220 | 441 |
| 6.3 | 2.94E-16 | 2.05E-17 | 14 | 221 | 441 |
| 11.2 | 1.05E-17 | 3.85E-19 | 27 | 223 | 421 |
| 15.2 | 6.20E-19 | 5.69E-20 | 11 | 224 | 436 |
| 24.5 | 3.81E-21 | 2.98E-20 | 0.13 | 211 | 437 |
| 35.0 | 2.77E-21 | 9.89E-21 | 0.28 | 223 | 443 |

## B: Temperature evolution of the β and 2β oscillations amplitudes

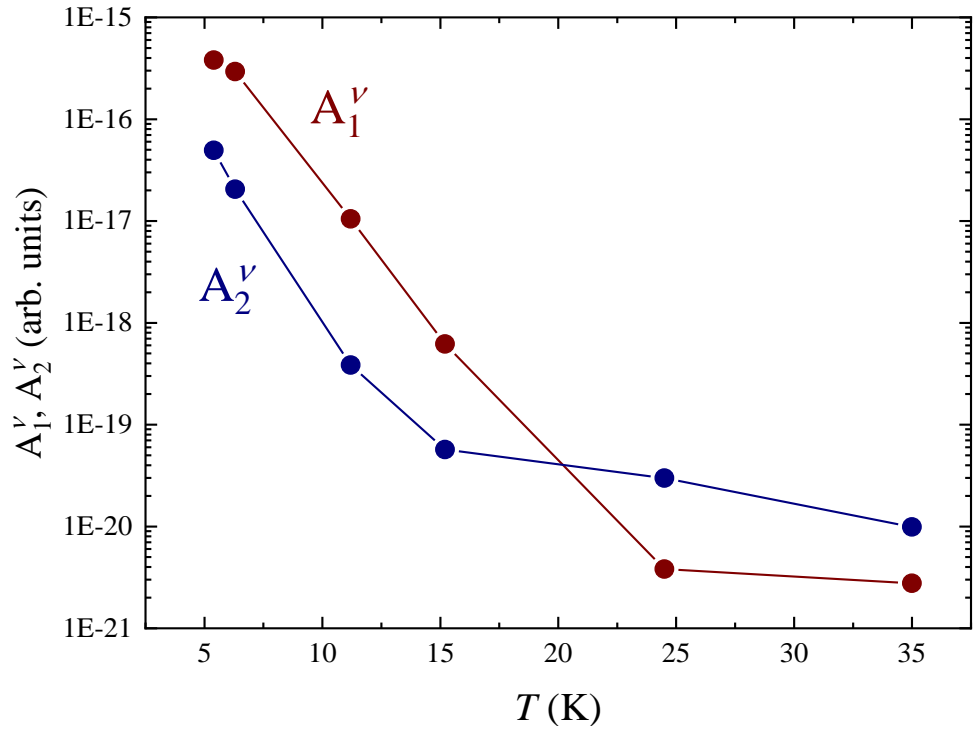

**Figure S2.** Calculated from the Nernst signal temperature dependence of the first ($A_1^\nu$) and second ($A_2^\nu$) harmonic amplitudes for the β band.

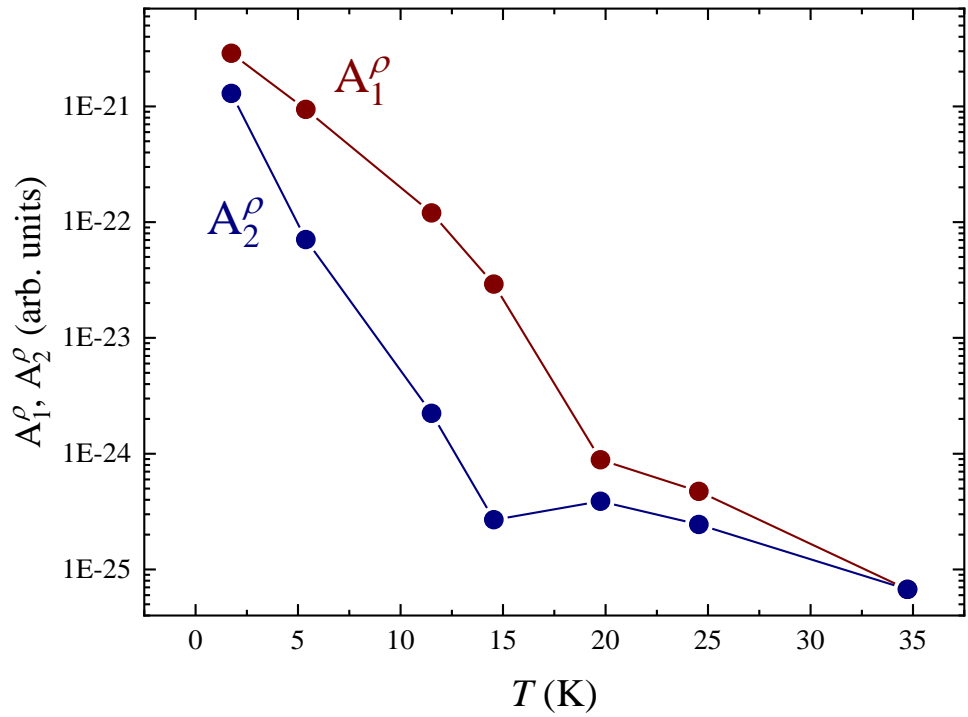

**Figure S3.** Calculated from the resistivity data temperature dependence of the first ($A_1^\rho$) and second ($A_2^\rho$) harmonic amplitudes for the β band.

**B: Filtered Nernst signal**

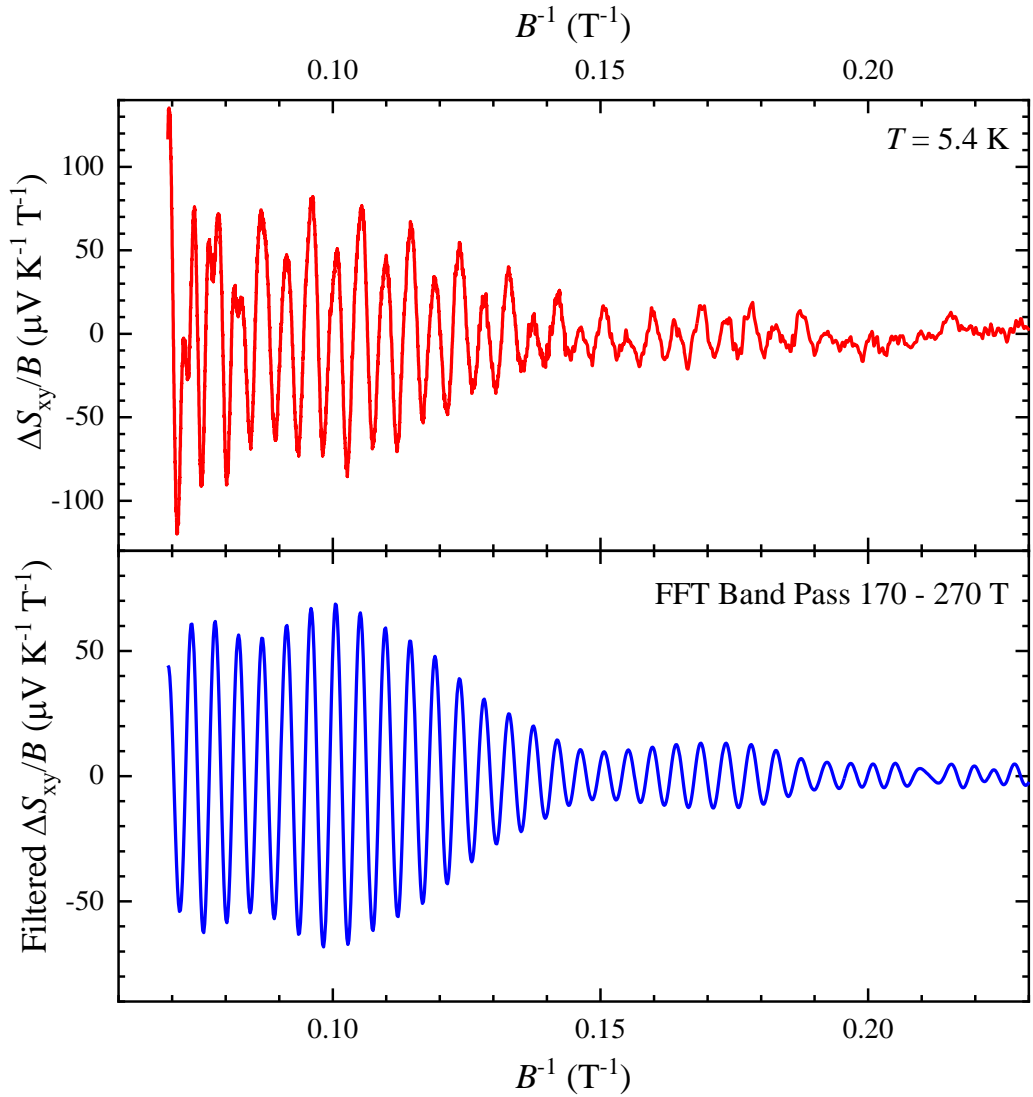

**Figure S4**. Upper panel: oscillatory Nernst signal in TaAs$_2$ measured at the temperature $T = 5.4$ K plotted versus inverse magnetic field. Lower panel: signal from the upper panel filtered with the FFT band pass filter for the 170 - 270 T frequency range (β oscillations).

## C: Effective mass calculation

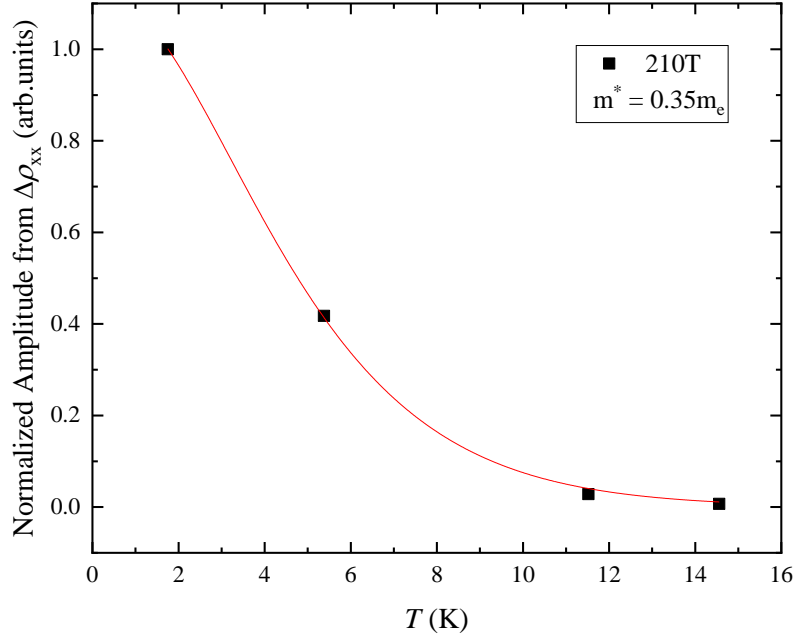

**Figure S5.** Temperature dependence of normalized FFT amplitude of the resistivity oscillations (8 T < $B$ < 14.5 T) for the β frequency in TaAs$_2$. Solid line is fit of the Lifshitz-Kosevich formula.

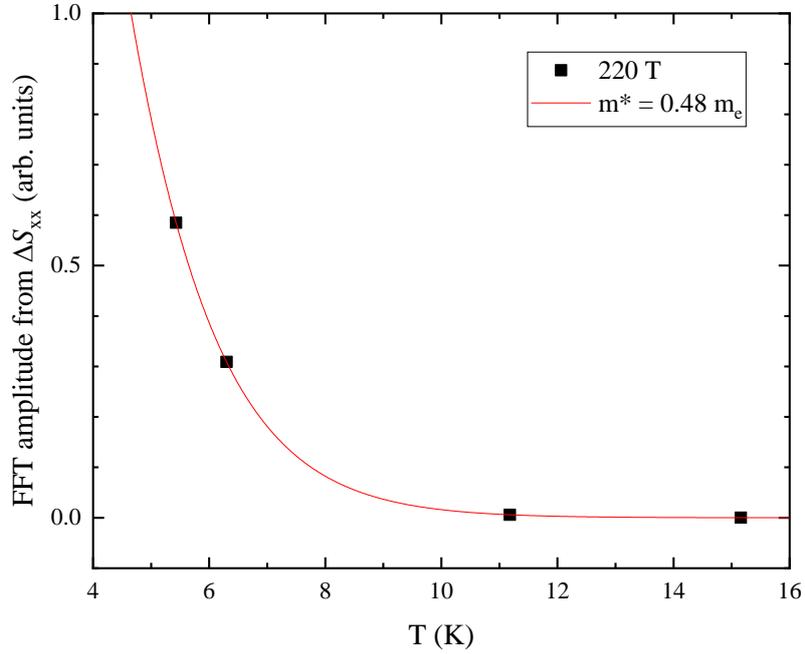

**Figure S6.** Temperature dependence of normalized FFT amplitude of the thermopower oscillations for the β frequency in TaAs$_2$, 5 T < B < 14.5 T. Solid line is fit of the modified Lifshitz-Kosevich formula [A.P. Morales et al., Phys. Rev. B **93**, 155120 (2016)]: $A(T) \propto \frac{(\alpha pX)\coth(\alpha pX)-1}{\sinh(\alpha pX)}$, where $\alpha = 2\pi^2 k_B/e\hbar$, $k_B$ is the Boltzmann constant, $e$ is the elementary charge, $\hbar$ is the reduced Planck constant, $p$ is the harmonic number, and $X = m^*T/B$.

## D: Amplitude ratio

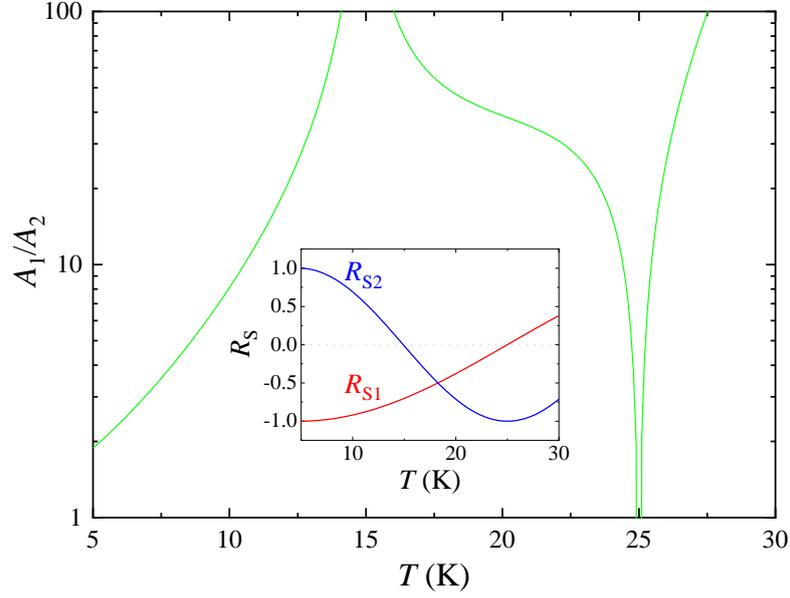

**Figure S7.** Temperature dependence of the absolute value of the first to second harmonic ratio calculated from the Lifshitz-Kosevich formula. The Dingle damping factor was assumed to be $R_D = 1$, the effective mass $m^* = 0.18\,m_0$, and the Landé g-factor linearly increasing with temperature to reach the spin zero condition at $T = 25$ K. Inset presents the temperature dependences of the spin damping factors for the fundamental frequency and its second harmonic. The latter goes through zero at $T \approx 15$ K which causes the second harmonic to disappear at this temperature. The resulting $A_1/A_2(T)$ qualitatively matches the experimental results, namely it initially (up to $T \approx 15$ K) increases, then decreases and goes through zero at $T = 25$ K and then increases again.

**E: Sample**

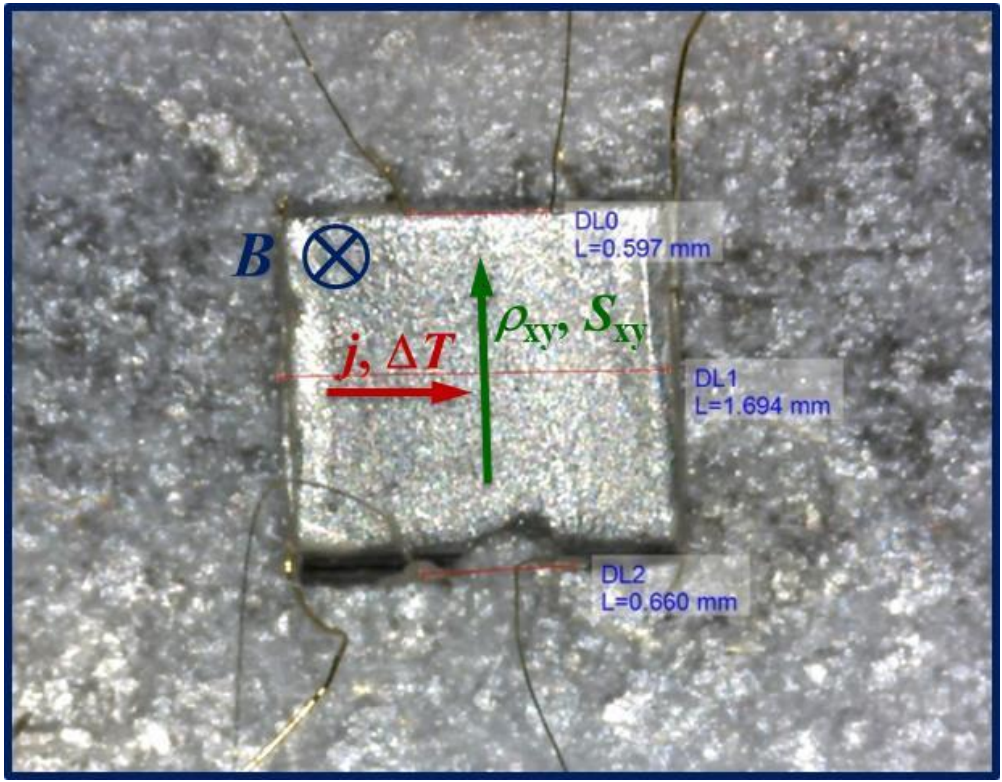

**Figure S8.** Single crystal of TaAs$_2$ with schematically shown experimental configuration.